\documentclass[conference]{IEEEtran}

\ifCLASSINFOpdf

\else

\fi

\usepackage{graphicx}

\usepackage{cite}
\usepackage{url}
\usepackage{hyperref}
\usepackage{filecontents}

\usepackage[cmex10]{amsmath}

\usepackage{algorithmic}
\usepackage{epstopdf}

\usepackage{array}
\usepackage{algorithmic}
\usepackage[ruled,norelsize]{algorithm2e}
\usepackage{algorithmic,algorithm2e,float}
\usepackage{multirow}
\usepackage[english]{babel}
\usepackage{amsmath}
\usepackage{amsthm}

\usepackage[tight,footnotesize]{subfigure}
\usepackage{color,soul}

\usepackage{mdwmath}
\usepackage{mdwtab}
\usepackage[top=0.9in, bottom=1.4in, left=0.63in, right=0.63in]{geometry}
\title{A Reconfigurable High-Performance Optical Data Center Architecture }

\author{\IEEEauthorblockN{Chong Liu, Maotong Xu, and Suresh Subramaniam }
\IEEEauthorblockA{Department of Electrical and Computer Engineering, The George Washington University\\
Email: cliu15@gwu.edu,  htfy8927@gwu.edu,  suresh@gwu.edu }}

\begin{document}
\maketitle


%
\IEEEpeerreviewmaketitle

\begin{abstract}
Optical data center network architectures are becoming attractive because of their low energy consumption, large bandwidth, and low cabling complexity. In~\cite{Xu1605:PODCA},
an AWGR-based passive optical data center architecture (PODCA) is presented. Compared with other optical data center architectures, e.g., DOS \cite{ye2010scalable}, Proteus \cite{singla2010proteus}, and Petabit \cite{xia2010petabit}, PODCA can save up to 90$\%$ on power consumption and 88$\%$ in cost. Also, average latency can be low as 9 $\mu$s at close to 100$\%$ throughput. However, PODCA is not reconfigurable and cannot optimize the network topology to dynamic traffic.

In this paper, we present a novel, scalable and flexible  reconfigurable architecture called RODCA. RODCA is built on and augments PODCA with a flexible localized intra-cluster optical network. With the reconfigurable intra-cluster network, racks with mutually large traffic can be located within the same cluster, and share the large bandwidth of the intra-cluster network. We present an algorithm for DCN topology reconfiguration, and present simulation results to demonstrate the effectiveness of reconfiguration.
\end{abstract}



\section{Introduction}
\label{sec:intro}
The number of data-intensive applications is rapidly increasing in data center networks. These applications, such as MapReduce, Hadoop, and Dropbox, require low latencies and high throughput and bring new challenges for future data center networks (DCNs). Data-intensive computing platforms typically use high speed communications switches and networks, which allow the data to be partitioned among the available computing resources and processed independently to achieve performance and scalability based on the amount of data.

In order to address these problems, recent research has focused on novel interconnect topologies for data center networks. The typical design is to place 20-50 servers in a rack, with an aggregation (``Top of Rack", or ToR) switch in each rack. Further, only a few ToRs are hot in DCNs and most of the traffic from these ToRs goes to a few other ToRs~\cite{kandula2009flyways},
requiring DCNs to be rapidly reconfigurable. Conventional DCNs, such as fat tree~\cite{al2008scalable}, flattened butterfly~\cite{kim2007flattened}, and VL2  \cite{greenberg2009vl2}, use commodity electrical switches to optimize the limited bandwidth available. However, these DCNs require a large number of links and switches, thus leading to rapidly increasing wiring complexity as the network scales. Moreover, electrical switches are also power-hungry devices. By contrast, optical DCNs provide the advantage of reduced power consumption and network cost as the network can be constructed using predominantly passive components~\cite{Kachris2012, chen2015optical, Xu1605:PODCA}.

\subsection{Related Work}
Existing optical data center networks are commonly based on optical switching, e.g., Semiconductor Optical Amplifier (SOA)-based switch, Micro-Electro-Mechanical Systems (MEMS) switches and Arrayed Waveguide Grating Routers (AWGR). MEMS switch, used in c-through \cite{wang2011c} and Helios \cite{farrington2011helios}, is a power-driven reconfigurable optical switch and its reconfiguration time could be on the order of a few milliseconds \cite{singla2010proteus}, and is therefore not well suited for fast packet switching in DCN applications. Nevertheless, optical MEMS switches could be reconfigured at coarse time scales for switching large volumes of data. Dynamically provisioning lightpaths for relatively stable traffic between server racks makes optical networks a cost-effective solution for allocating large bandwidth on-demand across the data center. Also, high-speed MEMS switches with switching times on the order of microseconds \cite{Kachris2012, porter2013integrating} are on the horizon, and are expected to be commercialized in the near future. Despite the relatively slow switching times, MEMS switches are scalable.
FARBON \cite{guo2014augmenting}, RODA \cite{pal2015roda} and Wavecube \cite{chen2015wavecube} are three recent reconfigurable optical DCN architectures that use specialized fast switches or expensive wavelength selective switches that do not scale well.

AWGR is a passive optical device that does not require reconfiguration, and can achieve packet contention resolution in the wavelength domain. The cyclic routing characteristic of the AWGR allows different inputs to reach the same output simultaneously by using different wavelengths.  Recently, a few AWGR-based DCN architectures have appeared in the literature such as DOS and Petabit. They employ tunable wavelength converters (TWCs), which are power-hungry devices \cite{Kachris2012}. Moreover, TWCs significantly increase the total cost of the architectures. We presented our own AWGR-based passive optical data center architecture (PODCA) in~\cite{Xu1605:PODCA}. There are three versions of the PODCA architecture suited to small (S), medium (M), and large (L) DCNs. Compared with DOS and Petabit, PODCA-L employs a large-scale AWGR as a central switch, and can easily accommodate over 2 million servers. We also presented algorithms for wavelength assignment and for scheduling packets in PODCA in~\cite{Xu1605:PODCA}. We showed in~\cite{Xu1605:PODCA} that average packet latencies (excluding protocol overhead) can be as low as 9 $\mu s$ at close to 100\% throughput. However, PODCA is not reconfigurable, and cannot adapt to fluctuations in traffic that are common in DCNs today.

\subsection{Our Approach and Contributions}
In this paper, we concurrently employ AWGR and MEMS-based optical switches to develop a scalable and reconfigurable architecture called RODCA. RODCA augments PODCA-L with a flexible intra-cluster optical network, which is a Clos multi-stage MEMS-based network, to adapt the network to traffic dynamics. The backbone of RODCA is a hierarchical optical DCN topology -- several ToRs are interconnected through an AWGR to form a {\em cluster}, and several clusters can be interconnected through a higher-level AWGR. The design for the intra-cluster network includes a reconfigurable switching network that can be reconfigured at relatively coarse time scales (i.e., reconfiguration times are comparable to or larger than packet transmission times), so that racks with mutually large traffic can be located within the same cluster, and enjoy the large bandwidth of the intra-cluster network.



The following contributions are made in this paper:
\begin{itemize}
\item We employ passive AWGRs and MEMS switches to develop a flexible and hierarchical DCN architecture.
\item We present an algorithm to trigger cluster reconfiguration based on traffic fluctuations.
\item We present extensive performance results from simulations exploring the effects of various algorithm and component parameters.
\end{itemize}

The rest of this paper is organized as follows: Section~\ref{sec:background} presents a brief description of the PODCA architecture. Section~\ref{sec:RODCA}  presents the proposed reconfigurable optical data center architecture, and an algorithm for topology reconfiguration in response to dynamic traffic changes. In Section \ref{sec:eval}, we present performance evaluation results.  Finally, we conclude the paper in Section~\ref{sec:conc}.

\begin{figure}[hb]
  \centering
  \label{fig:PODCA}
  \includegraphics[width=0.4\textwidth]{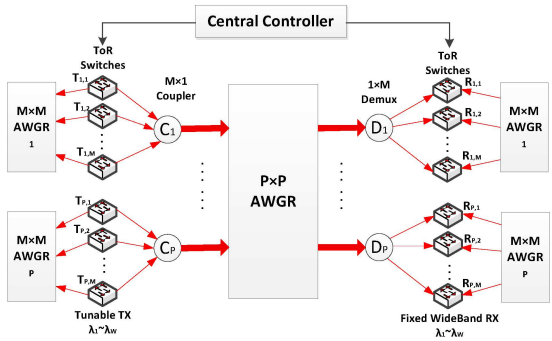}
  \caption{The PODCA-L architecture~\cite{Xu1605:PODCA}.}
  \vspace{-5pt}
\end{figure}


\section{Background}
\label{sec:background}
In this section, we briefly review PODCA-L~\cite{Xu1605:PODCA}, which is the architecture RODCA builds on. As shown in Fig. 1, PODCA-L is a hierarchical, AWGR-based architecture. Suppose the total number of racks is $S$ and a $P \times P$ AWGR is available. We split the $S$ racks into $P$ clusters and each cluster has $M$ racks, where $M = \left\lceil \frac{S}{P} \right\rceil$. $M$ racks of the same cluster connect to an input port of the $P\times P$ AWGR through a $M \times 1$ coupler and connect to an output port of the AWGR through a $1 \times M$ demultiplexer. The signal from an output port of a demultiplexer can be either a fixed wavelength or a fixed range of wavelengths. The $P \times P$ AWGR is for inter-cluster communication. Within each cluster, there is an $M \times M$ AWGR for intra-cluster communication. We denote $W = F \cdot P$ as the number of wavelengths, where $F \geq 1$ is an integer. The AWGR routes wavelengths from an input port to a specific output port in a cyclic way; the $c^{th}$ wavelength ${\lambda}_c$ is routed from input port \(i\) to output port \cite{Kachris2012}:
\begin{equation}\label{case_0}
  [(i+c-2) \mod \ P]+1, \ 1 \ {\leq} \ i{\leq} \ P, \ 1 \ {\leq} \ c \ {\leq} \ W.
\end{equation}

Each ToR has one or more fast tunable transmitters and fixed wide-band receivers. PODCA is a time-slotted system, where the time to transmit a packet is one time slot. Packets arriving to a ToR and needing to be transmitted to another ToR are placed in a virtual buffer in the ToR (one for each destination ToR). In each time slot, a central controller schedules packet transmissions for the next time slot and follows three scheduling constraints. First, a tunable transmitter or receiver can only transmit (respectively, receive) one packet at a time. Second, because of the cyclic wavelength routing property of the AWGR, at most $F$ packets can be transmitted from an input port of the AWGR to an output port of the AWGR in a time slot. Third, tunable transmitters connecting to the same AWGR port need to transmit on distinct wavelengths. The central controller first selects packets for transmission based on these three scheduling constraints, and then uses a packet scheduling algorithm to tune wavelengths and schedule packets for transmission.

In PODCA-L, some packets may need two hops to arrive at their destinations -- packets may need to be first routed to a ToR different from the destination ToR but in the same cluster as the destination ToR, by inter-cluster transmission, and then utilize intra-cluster transmission to reach the destination ToR. For each selected packet, the packet scheduling algorithm first checks if there is any available wavelength for transmitting directly to the packet's destination. If there is more than one available wavelength, a round robin method is used to choose one of the available wavelengths. If direct transmission is not possible, then the algorithm checks if a two-hop transmission is possible on any available wavelength. If this is not possible either, the packet waits in the source buffer until the next slot when the above steps are repeated. The interested reader can refer to~\cite{Xu1605:PODCA} for full details of the architecture and the scheduling algorithm.

\begin{figure*}[!t]
\centering
\includegraphics[height=2.0in,width=6.2in]{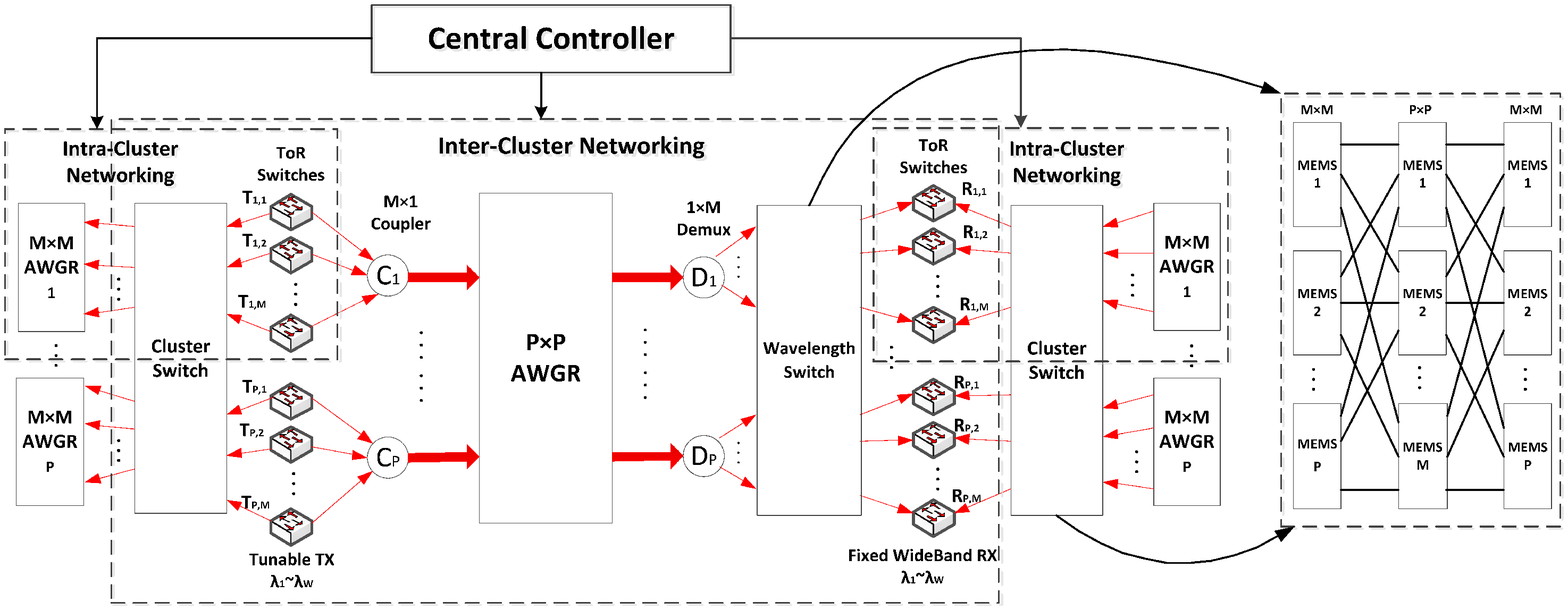}
\caption{The reconfigurable optical data center architecture (RODCA).}
\label{fig:RODCA}
\vspace{-10pt}
\label{fig_3}
\end{figure*}
\section{RODCA Architecture and Reconfiguration}
\label{sec:RODCA}
In this section, we present the proposed reconfigurable optical data center architecture (RODCA), and an algorithm for reconfiguration. Fig.~\ref{fig:RODCA} shows the proposed architecture.

\subsection{RODCA Design }
The architecture, shown in Fig.~\ref{fig:RODCA}, augments PODCA with two switches (switching networks, in fact) -- a {\em wavelength switch} (this is not to be confused with a wavelength-selective switch) and a pair of {\em cluster switches}. The cluster switches are used to reconfigure cluster memberships dynamically, i.e., to partition the set of rack transmitters and receivers into clusters based on traffic demands so that rack pairs with large traffic demands are placed in the same cluster. The use of the wavelength switch is explained below.

Currently, the size of the fast (microsecond-level) MEMS optical switch mentioned in \cite{Kachris2012} is limited to a few tens of ports. In order to scale the network to large sizes (hundreds or thousands of racks), we use a Clos multi-stage network of MEMS optical switches to build large-sized cluster and wavelength switches. The path diversity and non-blocking nature of the Clos network enables the routing of arbitrary traffic patterns with no loss of throughput \cite{kim2007flattened}, and plays an important role in the scalability of RODCA.

We now explain why the wavelength switch is needed. During the topology reconfiguration process, RODCA reconfigures the wavelength switch to make each cluster to be able to receive all $W$ wavelengths from an output port of the inter-cluster AWGR. Without the wavelength switch, some racks might not be able to receive certain wavelengths. To illustrate this point, in Fig. 3, consider a network of four racks, $R_{1,1}$, $R_{1,2}$, $R_{2,1}$, and $R_{2,2}$, and suppose two wavelengths are available. Without the wavelength switch, $R_{1,1}$ and $R_{2,1}$ can only receive $\lambda_1$, and $R_{1,2}$ and $R_{2,2}$ can only receive $\lambda_2$ (as per the design of PODCA~\cite{Xu1605:PODCA}, which uses fixed-band receivers). After reconfiguration, suppose $R_{1,1}$ and $R_{2,2}$ are placed into one cluster. Then, if the wavelength switch is not used, this cluster cannot receive any packet from racks that are connected to the second/bottom input port of the inter-cluster AWGR, since the second input port of the inter-cluster AWGR can reach the first/top output port of the inter-cluster AWGR only by using $\lambda_2$, but this wavelength cannot be received by $R_{1,1}$. Similarly, the second  input port of the inter-cluster AWGR can reach the second output port of the inter-cluster AWGR  by using $\lambda_1$, but the wavelength cannot be received by $R_{2,2}$. However, by using the wavelength switch, we can connect $R_{1,1}$ and $R_{2,2}$ to the second  output port of the inter-cluster AWGR. Now, $R_{1,1}$ and $R_{2,2}$ can receive $\lambda_1$ and $\lambda_2$, respectively. The second input port of the inter-cluster AWGR can reach this cluster by sending packets to $R_{1,1}$ by using $\lambda_1$; it can also reach it by sending packets to $R_{2,2}$ on $\lambda_2$.
\begin{figure}[hb]
  \centering
  \label{wavelength_switch}
  \includegraphics[height=1.3in,width=3.3in]{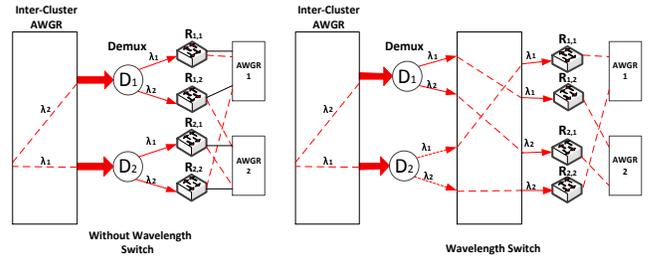}
  \caption{The need for the wavelength switch.}
  \vspace{-10pt}
\end{figure}

\subsection{Reconfiguration Algorithm}
We use the number of packets in the virtual buffers to decide {\em when} the RODCA topology should be reconfigured. In particular,
we consider the total number of packets waiting for inter-cluster transmission ($L_{inter}$), the total number of packets waiting for intra-cluster transmission ($L_{intra}$), and use a threshold parameter ($\beta$) to make the reconfiguration decision. For this purpose, every rack is required to send the number of packets waiting in each of its virtual buffers to the central controller. We then propose to perform reconfiguration if:
\begin{equation}\label{con_1}
  {\beta} \ {\cdot} \ {L_{\rm intra}} \ {\leq} \ {L_{\rm inter}},
\end{equation}
Thus, reconfiguration is triggered if (a multiple of) the number of intra-cluster packets falls below the number of inter-cluster packets. Since sampling the buffers and reporting the buffer occupancy to the controller causes control message overhead, we introduce a sampling interval (SI) to balance the overhead with performance. SI = $\delta$ means that the buffers are sampled every $\delta$ time slots, and the reconfiguration condition above is checked.

If the condition is satisfied, the controller suspends all inter- and intra-cluster packet transmissions in order to reconfigure the network based on the number of packets waiting for transmission in the virtual buffers. We denote a set containing all $P{\cdot}M$ racks as $\Theta$ and call the number of packets between two racks as the {\em mutual number of packets}. A simple greedy heuristic is used to form clusters. We first find two racks with the largest mutual number of packets in $\Theta$, and remove these two racks from $\Theta$. Then, we find another rack in $\Theta$ with the largest mutual number of packets with the previous two chosen racks and remove this rack from $\Theta$. We repetitively find a rack from $\Theta$ owning the largest mutual number of packets with all chosen racks in the previous two steps and remove the rack from $\Theta$, and so on, until we have collected $M$ racks to generate a cluster. We repeat the above steps until we have $P$ clusters. The pseudocode of this reconfiguration algorithm, which is executed every time slot, is shown in Algorithm \ref{alg2}. In Algorithm \ref{alg2}, we use a counter to indicate if it is time to reconfigure.

\begin{algorithm}                      
\algsetup{linenosize=\small}
  \small
\caption{\sc Reconfig}          
\label{alg2}                           
\begin{algorithmic}[1]               
\IF {counter == SI and $\beta$ $\cdot$ $L_{\rm intra}$ $\leq$ $L_{\rm inter}$}
\STATE{counter = 0}
\FOR {$p=1:P$}
    \STATE {find a pair of racks with the largest mutual number of packets}
    \STATE {remove this pair of racks from $\Theta$}
    \FOR {$m=1:M$}
        \STATE {find a rack with the largest mutual number of packets for all racks removed in iteration $p$}
        \STATE {remove the rack from $\Theta$}
    \ENDFOR
\ENDFOR
\ELSE
\STATE {counter++}
\ENDIF
\end{algorithmic}
\end{algorithm}

Once reconfiguration is complete, we resume packet transmission by following the {\em packet scheduling for PODCA-L} algorithm~\cite{Xu1605:PODCA}. The central controller schedules both intra-cluster and inter-cluster transmission based on {\em scheduling constraints}. Recall that some packets can be transmitted in one hop, and some need two hops to arrive at their destinations. For example, suppose \(P\) equals \(2\) and \(W\) equals \(8\). There are \(8\) racks within each cluster, i.e., $M = 8$. On each rack, there is one tunable transmitter and one wide-band receiver for intra-cluster communication. Also, there is one tunable transmitter and one wide-band receiver for inter-cluster communication. Suppose a packet is from $T^{1}_{1,1}$ to $R^{1}_{2,1}$. The only wavelength $R^{1}_{2,1}$ can receive, within inter-cluster transmission, is ${\lambda}_1$. However, based on the routing characteristics of the AWGR, the receivable wavelengths from the first input port of the AWGR to the second output port of the AWGR can only be ${\lambda}_2$, ${\lambda}_4$, ${\lambda}_6$ and ${\lambda}_8$. Therefore, ${\lambda}_1$ transmitted from $T^{1}_{1,1}$ cannot arrive at the second output port of the AWGR in a single hop. Thus, a two-hop transmission is needed. We first choose one wavelength from $\{$${\lambda}_2$, ${\lambda}_4$, ${\lambda}_6$, ${\lambda}_8$$\}$; suppose we choose ${\lambda}_2$. The packet is transmitted to $R_{2,2}$ by using ${\lambda}_2$, and then in the next time slot, $R_{2,2}$ can transmit the packet to $R_{2,1}$ by using an intra-cluster transmission.

\begin{algorithm}                   
\algsetup{linenosize=\small}
  \small
\caption{\sc RODCA\_Packet\_Scheduling}          
\label{alg3}                           
\begin{algorithmic}[1]                    
\FOR {packets at the head of virtual buffers}
    \STATE {\sc Reconfig ()}
    \STATE {Load-balance between inter- and intra-cluster network by using {\sc PODCA-L\_Packet\_Scheduling ()}} \cite{Xu1605:PODCA}
\ENDFOR
\end{algorithmic}
\end{algorithm}

If the source and destination racks are in the same cluster, we can use either intra-cluster transmission or inter-cluster transmission. Here, we define a threshold to determine if the packet uses intra-cluster transmission or inter-cluster transmission. If the number of packets waiting for intra-cluster transmission is less than the threshold, then we place the packet at the tail of the waiting queue of the intra-cluster transmission. Otherwise, we use inter-cluster transmission to transmit that packet. To transmit more than one packet in a time slot, each ToR can have more than one tunable transmitter and wide-band receiver for intra-cluster transmission, or inter-cluster transmission, or both.

The operation of RODCA follows the algorithm whose pseudocode is shown in Algorithm \ref{alg3}.

\section{Performance Evaluation}
\label{sec:eval}
In this section, we conduct simulations to evaluate the latency and throughput performance of RODCA. For space reasons, we fix the network configuration and conduct experiments by varying the algorithm and component parameters only. Both the inter-cluster and intra-cluster AWGR sizes are set to be $30{\times}30$; thus, the DCN is made up of 900 racks, and we assume $48$ servers per rack for a total of 43,200 servers in the DCN. The number of wavelengths is set to 120. On each rack, there are 3 tunable transmitters for intra-cluster transmission and 1 tunable transmitter for inter-cluster transmission. Each rack has $P{\cdot}M=900$ virtual queues and each virtual queue buffers packets for each destination rack. The buffer size of a ToR is 5 Mb. The transmission rate of each tunable transmitter is assumed to be 10 Gbps, and its tuning time is $8$ ns~\cite{Xu1605:PODCA}. Since the network only connects racks with each other, we assume traffic arrives to racks, and do not model server traffic in our simulations. Following~\cite{guo2014augmenting}, we assume that tasks arrive to the data center according to a Poisson process with a mean rate of 3 tasks/sec. Each task arrives to a random source rack, and triggers $\kappa$ flows, where $\kappa \in [1, 900]$ is an integer chosen based on a uniform distribution. We model two types of traffic flows, {\em mice flows} and {\em elephant flows}. Following~\cite{Srikanth2009}, we use two Gaussian distributions to model mice and elephant flow rates. (If the Gaussian distribution gives a negative value, we set the value to be 0.) Mice flow arrival rates for each server follow a Gaussian distribution with mean $0.01$ Mbps, and elephant flow arrival rates for each server follow a Gaussian distribution with mean 40 Mbps. Thus, mice arrival rates for each ToR follow a Gaussian distribution with mean 0.48 Mbps ($48 \times 0.01$ Mbps), and elephant flow arrival rates for each ToR follow a Gaussian distribution with mean 1.92 Gbps ($48 \times 40$ Mbps). We dynamically change the number of elephant flows and mice flows to make the ratio between total elephant arrival rate and total mice arrival rate to be 9:1.

Recall that $\beta$ is a threshold parameter that is used by our algorithm to trigger network reconfiguration. Each packet is assumed to be 1500 bytes long, which implies that each slot is $1.2 \ \mu s$ long (assuming $10$ Gbps wavelength capacity and transmitter rates). The time to reconfigure the topology, RT, is set to 1 slot (fast switches) or 10 slots (relatively slow switches). In all of our simulations, we observed no packet drops and thus the throughput is $100\%$, except for the case when $RT = 10$ and $\beta = 0$. In this case, we observed a $5\%$ packet drop rate, suggesting that too frequent reconfiguration can be detrimental to performance when RT is not small.\footnote{We will see later that the latency is also negatively impacted in this scenario.} We therefore focus on packet latency, and make references to throughput only when it is less than $100\%$.

\begin{figure}[h]
\centering
\includegraphics[height=1.40in,width=0.35\textwidth]{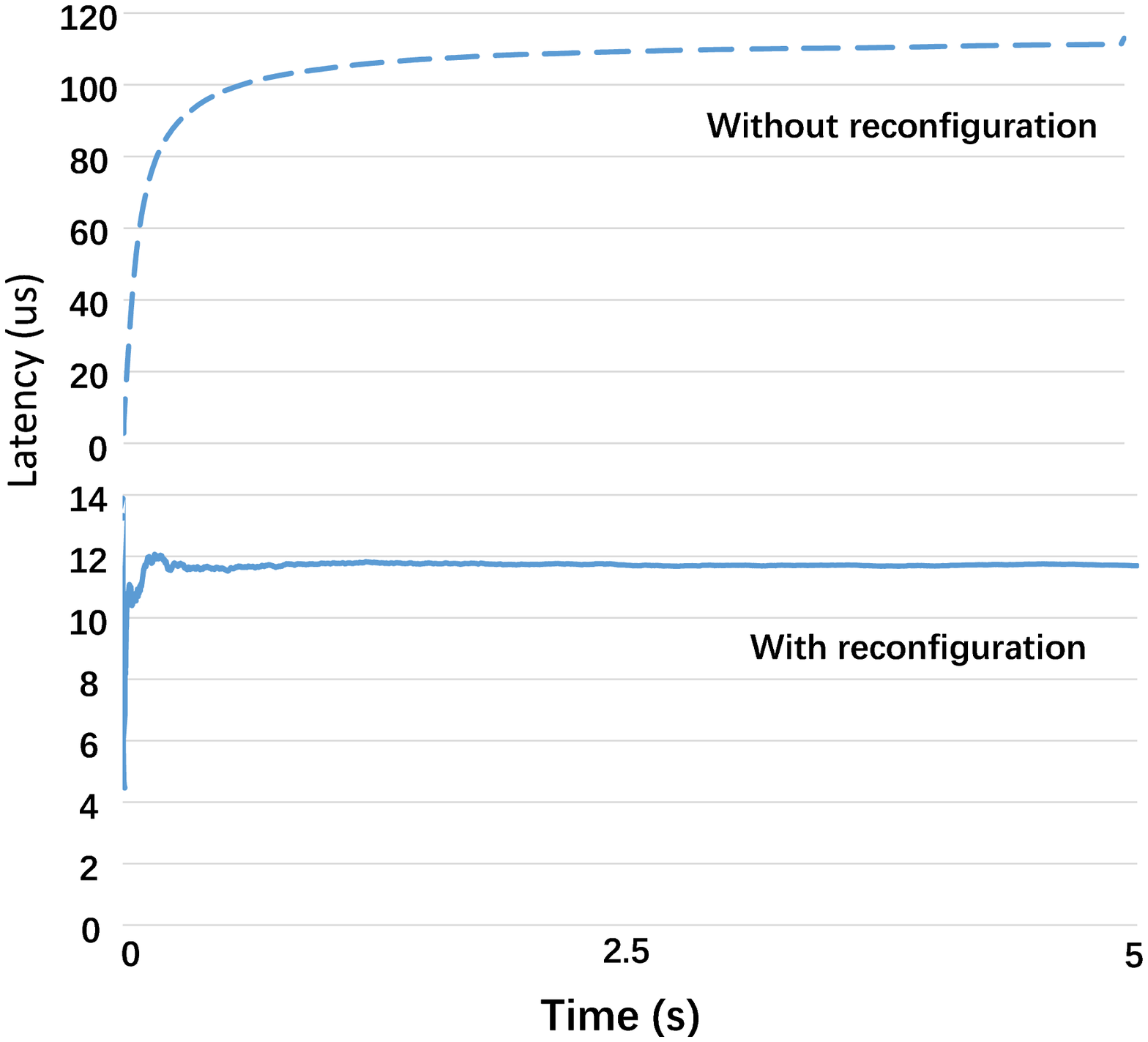}
\caption{The effect of reconfiguration on packet latency.}
\label{Fig_e_1}
\vspace{-6pt}
\end{figure}

We first compare the throughput and latency performance with and without reconfiguration. We set SI = 1, $\beta =15$, and RT = 10, and simulate the DCN for 5 seconds (corresponding to approximately 600 million packet arrivals). Our results show that no packets are dropped with reconfiguration, whereas approximately $4\%$ of the packets are dropped when the network does not reconfigure in response to traffic changes. Fig.~\ref{Fig_e_1} shows the evolution of packet latency with time as the network starts from an empty state. The latency quickly reaches steady state and settles around 12 $\mu$s with reconfiguration, whereas it increases rapidly to around 110 $\mu$s when no reconfiguration is done. This 90$\%$ improvement in latency clearly shows the benefits of adapting the network topology to dynamic traffic variations.

\begin{figure}[h]
\centering
\includegraphics[height=1.30in,width=0.35\textwidth]{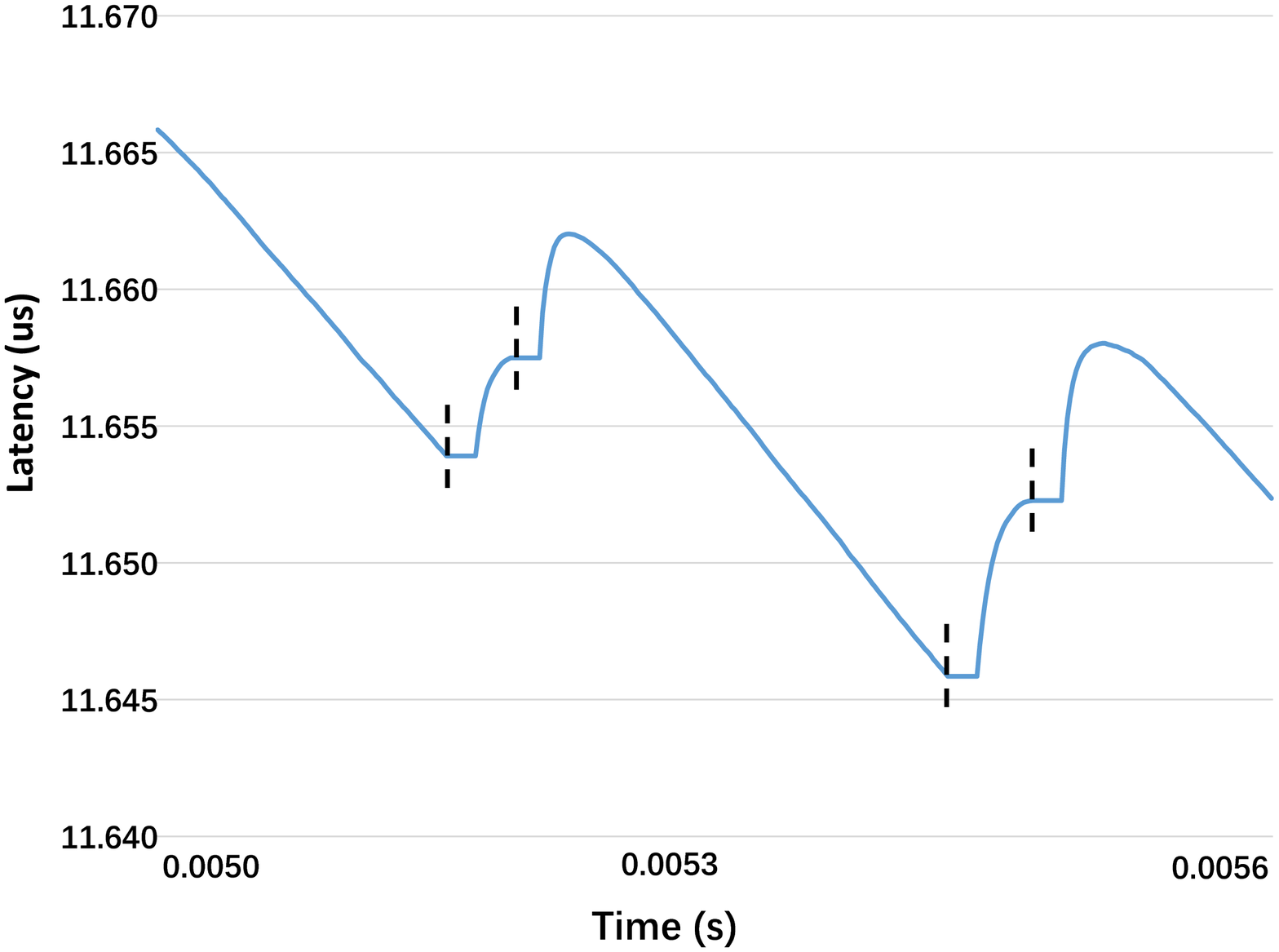}
\caption{Latency dynamics upon reconfiguration.}
\label{Fig_e_2}
\vspace{-7pt}
\end{figure}

In Fig.~\ref{Fig_e_2}, we zoom into a 0.6 ms window of Fig.~\ref{Fig_e_1} to better illustrate the behavior of latency when reconfiguration happens. Reconfiguration initiation times are marked with a dashed vertical line in the figure. When the reconfiguration condition is satisfied, RODCA initiates reconfiguration, and suspends all packet transmissions. During this time, no packets are received by any destination rack and thus the packet latency remains constant. RODCA resumes packet transmission after reconfiguration is complete, and packet latency starts to increase initially and then goes down. The initial increase in latency is due to the fact that packets that are backlogged during reconfiguration contribute to an increase in latency, but soon after, the latency starts decreasing because of the optimized topology. Fig.~\ref{Fig_e_2} also shows that two reconfigurations could occur back to back, and the effect of the first reconfiguration on packet latency is never noticed. This is a consequence of choosing a small value of SI. In our experiments, we have not seen this occur when SI and $\beta$ are large.

\begin{figure}[h]
\centering
\includegraphics[height=1.6in,width=0.35\textwidth]{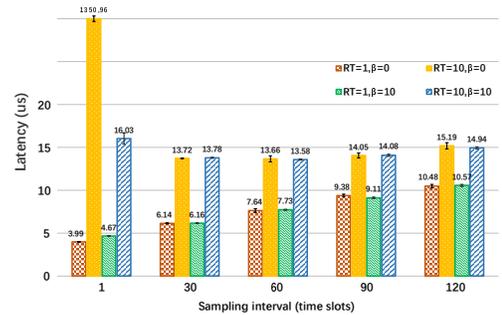}
\caption{Latency vs.\ sampling interval for various values of $\beta$ and RT.}
\label{Fig_e_3}
\vspace{-6pt}
\end{figure}

We next show the latency for different SI, $\beta$, and RT values in Fig.~\ref{Fig_e_3}. For this and subsequent figures, we simulate 100 million packet arrivals, and
show 95$\%$ confidence intervals from 30 trials for each data point.  Fig.~\ref{Fig_e_3} shows that when RT = 1, small SI and $\beta$ give the best latency performance. Intuitively, when RT is small, there is minimal penalty for reconfiguring the network (which is facilitated by small values of SI and $\beta$) and optimizing the topology helps boost performance significantly. Further, when RT = 10, as the SI increases, the latency first decreases and then increases. As the SI increases, the frequency of reconfiguration decreases, and as Fig.~\ref{Fig_e_2} illustrates, some reconfigurations are not useful at all when SI is small. Thus, when SI increases from 1 to 60, the penalty from suspension of packet transmissions decreases and the number of useless reconfigurations decreases. However, when SI increases from 60 to 120, the benefit from reconfigurations also decreases and the latency increases. Eventually, as SI approaches infinity, reconfiguration is rarely triggered, and the latency would approach that for the no reconfiguration case. Fig.~\ref{Fig_e_3} also illustrates how $\beta$ affects the packet latency. A large $\beta$ implies a reluctance to reconfigure the network, and this is particularly useful when SI is small, e.g., 1. It can be seen that a large $\beta$ tremendously decreases the packet latency in this case.

\begin{figure}[h]
\centering
\includegraphics[height=1.4in,width=0.35\textwidth]{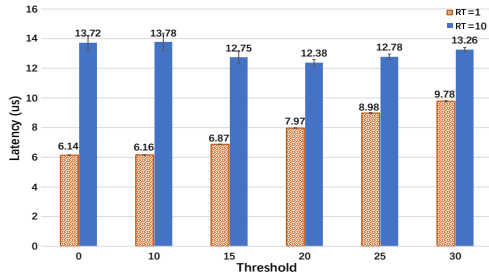}
\caption{Latency vs. threshold for various values of RT}
\label{Fig_e_4}
\vspace{-6pt}
\end{figure}

Fig.~\ref{Fig_e_4} presents packet latency for different values of the threshold $\beta$, for SI = 30. When RT = 1, as $\beta$ increases, the latency always increases. When RT = 10, as $\beta$ increases, the latency first decreases and then increases, as explained above. Fig.~\ref{Fig_e_4} shows that once an SI is chosen, we can tune the $\beta$ to get the smallest latency. Even though both SI and $\beta$ affect the latency, they are introduced for different purposes. Recall that SI is a measure of how frequently the virtual buffers are sampled.\footnote{We have also looked into averaging the sampled values over a time window, and found it to be not very useful.} This frequency places a load on the controller, with smaller SI values requiring more frequent control messages for gathering the occupancy information on the virtual buffers. The actual overhead due to such messages is not modeled in this paper, and is left for future work. In contrast to SI, $\beta$ is an algorithm parameter for controlling the latency for a given SI.


%
%

\section{Conclusions}
\label{sec:conc}
In this paper, we present a scalable and flexible reconfigurable architecture called RODCA. RODCA is built on and augments PODCA-L with a flexible intra-cluster optical network. With the reconfigurable intra-cluster network, racks with mutually large traffic can be located within the same cluster, and share the large bandwidth of the intra-cluster network. We present an algorithm for DCN topology reconfiguration, and present simulation results to demonstrate the effectiveness of reconfiguration. Our results show that packet latencies around 10-12 $\mu$s are achievable with judicious topology reconfiguration in response to dynamic traffic changes, while latencies could be an order of magnitude larger if the topology is not reconfigured.



\section{Acknowledgments}
\label{sec:ack}
This work was supported in part by NSF award \# 1618487.



%

\bibliographystyle{IEEEtran}
\bibliography{references}

\end{document}